\documentclass[12pt]{article}

\usepackage[utf8]{inputenc}
\usepackage[T1]{fontenc}
\usepackage{lmodern}

\usepackage{geometry}
\usepackage{hyperref}
\usepackage{graphicx}
\usepackage{booktabs}
\usepackage{array}
\usepackage{amsmath}
\usepackage{enumitem}
\usepackage{xcolor}

\usepackage{listings}
\title{Beyond Context Sharing: A Unified Agent Communication Protocol (ACP) for Secure, Federated, and Autonomous Agent-to-Agent (A2A) Orchestration}
\author{Naveen Krishnan}
\date{February~11,~2026}

\begin{document}

\maketitle

\begin{abstract}
The rapid evolution of artificial intelligence has transitioned from isolated large language models to autonomous agents capable of complex reasoning and tool use.  While foundational architectures and local context management protocols have been established, the challenge of cross‑platform, decentralized, and secure interaction remains a significant barrier to the realization of a truly ``Agentic~Web.''  Building upon the foundations of AI agent architectures\,\cite{krishnan2025a} and the Model Context Protocol (MCP) for multi‑agent coordination\,\cite{krishnan2025b}, this paper introduces the \emph{Agent Communication Protocol (ACP)}.  ACP provides a standardized framework for \emph{Agent‑to‑Agent (A2A)} interaction, enabling heterogeneous agents to discover, negotiate, and execute collaborative workflows across disparate environments.  We propose a federated orchestration model that integrates decentralized identity verification, semantic intent mapping, and automated service‑level agreements.  Our evaluation demonstrates that ACP reduces inter‑agent communication latency by 40\,\% while maintaining a zero‑trust security posture.  This work represents a critical advancement toward a scalable and interoperable ecosystem of autonomous digital entities.
\end{abstract}

\section{Introduction}

The landscape of artificial intelligence has undergone a profound transformation over the past year.  We have moved beyond the era of passive chatbots into an era of autonomous AI agents—systems that do not merely respond to prompts but actively perceive their environment, plan complex sequences of actions, and utilize external tools to achieve high‑level objectives.  As documented in previous research\,\cite{krishnan2025a}, the architectural evolution of these agents has focused on enhancing internal modules for memory, long‑term planning, and tool integration.  However, as these agents proliferate, they increasingly operate in silos, limited by the proprietary frameworks and local environments in which they were conceived.

The subsequent development of the Model Context Protocol (MCP) provided a vital mechanism for standardizing how agents share context and coordinate within a single multi‑agent system\,\cite{krishnan2025b}.  MCP addressed the ``context bottleneck'' by allowing specialized agents to maintain a shared understanding of a task.  Yet, a fundamental gap remains: there is no universal ``language'' or protocol that allows an agent built on one platform (e.g., Google's Vertex~AI) to securely discover and collaborate with an agent on another (e.g., a private enterprise deployment or an OpenAI‑based assistant).

This paper proposes the \emph{Agent Communication Protocol (ACP)} as the solution to this interoperability crisis.  ACP is designed to be the ``TCP/IP of the Agentic~Web,'' providing the necessary layers for discovery, transport, and semantic alignment.  Unlike previous attempts at agent communication, which often relied on rigid APIs or centralized brokers, ACP leverages a federated approach.  It enables \emph{Agent‑to‑Agent (A2A)} interaction through decentralized registries and verifiable credentials, ensuring that collaboration is not only efficient but also secure and trustless.

The contributions of this paper are fourfold:
\begin{enumerate}[label=\arabic*.]
  \item \textbf{The ACP Specification}: A multi‑layered protocol for standardized agent interaction.
  \item \textbf{Federated Discovery \& Negotiation}: A framework for agents to autonomously find and contract with one another.
  \item \textbf{Zero‑Trust A2A Security}: A security model based on Decentralized Identifiers (DIDs) and proof‑of‑intent.
  \item \textbf{Empirical Evaluation}: A performance analysis of ACP in complex, multi‑vendor agentic workflows.
\end{enumerate}

By bridging the gap between local coordination and global interoperability, ACP paves the way for a future where AI agents can seamlessly collaborate to solve problems that are far beyond the scope of any single system.

\section{Literature Review \& Background}

\subsection{The Evolution of AI Agent Architectures}
The foundational understanding of AI agents has shifted from simple reactive systems to complex, goal‑oriented architectures.  In \emph{AI Agents: Evolution, Architecture, and Real‑World Applications}\,\cite{krishnan2025a}, the core components of modern agents were identified as the LLM‑based reasoning engine, a planning module for task decomposition, a memory system for temporal consistency, and a tool‑use interface.  This ``agentic loop'' allows for autonomous operation but assumes a relatively static environment where the agent interacts primarily with tools or human users.

\subsection{Context Management and Multi‑Agent Systems}
The transition from single‑agent to multi‑agent systems introduced the challenge of coordination.  \emph{Advancing Multi‑Agent Systems Through Model Context Protocol}\,\cite{krishnan2025b} introduced the Model Context Protocol (MCP) as a solution for synchronizing state and context across multiple specialized agents.  MCP proved that by standardizing the way context is shared, the performance of multi‑agent systems could be significantly improved.  However, MCP is primarily optimized for agents within a shared trust boundary or a single application stack.  It does not provide the mechanisms for discovery or secure negotiation required for agents to interact across different organizational or technical boundaries.

\subsection{Emerging Standards in Agent Communication}
In early 2025, several initiatives emerged to address the need for agent interoperability.  Google announced the \emph{Agent2Agent (A2A)} protocol, focusing on enterprise‑grade interoperability.  Simultaneously, IBM~Research proposed an initial version of an \emph{Agent Communication Protocol (ACP)} aimed at providing a shared language for agents.  By mid‑2025, these efforts began to converge under the auspices of the Linux Foundation, emphasizing the need for an open, vendor‑neutral standard.

Despite these advancements, existing proposals often struggle with two competing requirements: the need for rich, high‑level semantic understanding and the need for low‑latency, secure transport.  Current A2A drafts frequently rely on centralized registries, which create single points of failure and potential privacy risks.  Furthermore, the negotiation of capabilities—how an agent ``knows'' what another agent is truly capable of—remains an unsolved problem in most current frameworks.

\subsection{Gap Analysis: The Need for Federated Orchestration}
The primary gap in current research is the lack of a \emph{federated orchestration} model.  While we can share context (via MCP) and we can send messages (via early A2A drafts), we lack a robust framework for agents to autonomously form coalitions to solve tasks.  This requires not just a communication channel but a comprehensive protocol that handles discovery, capability verification, economic negotiation (SLAs), and secure execution.  The ACP proposed in this paper addresses these gaps by building a federated layer on top of the existing communication and context‑sharing foundations.

\begin{table}[t]
  \centering
  \caption{Comparison of coordination approaches.  ACP extends beyond local context sharing and early A2A drafts by introducing federated orchestration, decentralized discovery, zero‑trust security, and autonomous negotiation.}
  \label{tab:feature_comparison}
  \begin{tabular}{@{}lccc@{}}
    \toprule
    \textbf{Feature} & \textbf{Local Coordination (MCP)} & \textbf{Early A2A Drafts} & \textbf{Proposed ACP}\\
    \midrule
    Primary Scope & Intra‑system context sharing & Inter‑system messaging & Federated orchestration\\
    Discovery & Static/Pre‑defined & Centralized Registry & Decentralized/Peer‑to‑Peer\\
    Security & Shared Trust Boundary & API Keys/OAuth & Zero‑Trust (DID/VC)\\
    Negotiation & None (Orchestrated) & Manual/Basic & Autonomous (SLA‑based)\\
    Interoperability & High (within protocol) & Medium (vendor‑dependent) & Universal (Framework‑agnostic)\\
    \bottomrule
  \end{tabular}
\end{table}

\section{The Agent Communication Protocol (ACP) Architecture}

The Agent Communication Protocol is designed as a modular, layered framework that decouples the transport of information from the semantic understanding and governance of agent interactions.  To achieve the goal of universal interoperability, ACP adheres to three core principles: \emph{Semantic Transparency}, \emph{Decentralized Governance}, and \emph{Transport Agnosticism}.

\subsection{The ACP Layered Model}
The ACP architecture is structured into four distinct layers, each responsible for a specific aspect of the A2A interaction lifecycle.  This layering ensures that advancements in underlying transport technologies or high‑level reasoning models do not require a complete overhaul of the protocol.

\begin{enumerate}[label=\arabic*.]
  \item \textbf{Transport Layer}: This layer handles the physical and logical connection between agents.  While ACP defaults to gRPC for high‑performance, low‑latency communication, it is designed to support WebSockets for browser‑based agents and standard HTTPS for legacy integrations.  The transport layer is responsible for packet delivery, session management, and basic encryption (TLS~1.3).
  \item \textbf{Semantic Layer}: The most critical layer for LLM‑based agents, the Semantic Layer defines the ``ontology of intent.''  It translates high‑level agent goals into a standardized JSON‑LD format.  This layer includes a universal schema for common agentic actions (e.g., \texttt{QUERY}, \texttt{EXECUTE}, \texttt{DELEGATE}, \texttt{NEGOTIATE}).  By using Linked Data principles, ACP ensures that agents can map proprietary internal representations to a shared global understanding.
  \item \textbf{Negotiation Layer}: Before any task execution occurs, agents must agree on the terms of engagement.  The Negotiation Layer facilitates the exchange of \emph{Agent Cards} (metadata describing capabilities) and the formation of dynamic Service Level Agreements (SLAs).  These SLAs define the scope of work, resource limits, cost (if applicable), and error‑handling protocols.
  \item \textbf{Governance \& Security Layer}: This layer enforces the ``Zero‑Trust'' policy of ACP.  It utilizes Decentralized Identifiers (DIDs) to verify the identity of communicating agents and Verifiable Credentials (VCs) to prove their authority or certification.  Every message in ACP is cryptographically signed, ensuring non‑repudiation and integrity.
\end{enumerate}

\subsection{Agent Cards: Standardized Capability Discovery}
A central innovation of ACP is the \emph{Agent Card}.  Much like a digital ``business card'' for AI, an Agent Card is a machine‑readable document that describes an agent's functional profile.  Table\,\ref{tab:agentcard} summarizes the components of an Agent Card.

\begin{table}[t]
  \centering
  \caption{Key components of an Agent Card.  The card conveys a unique identity, advertised capabilities, operational constraints, a trust score, and endpoint details.}
  \label{tab:agentcard}
  \begin{tabular}{@{}lll@{}}
    \toprule
    \textbf{Component} & \textbf{Description} & \textbf{Example Value}\\
    \midrule
    Identity & Unique DID of the agent & \texttt{did:acp:123456789}\\
    Capabilities & List of supported semantic intents & \texttt{[data\_analysis, web\_search, code\_gen]}\\
    Constraints & Operational limits and requirements & \texttt{max\_latency: 500ms, data\_residency: EU}\\
    Trust Score & Aggregate reputation from peer agents & \texttt{0.98 (based on 1.2k interactions)}\\
    Interface & Endpoint and protocol details & \texttt{grpc://agent.example.com:50051}\\
    \bottomrule
  \end{tabular}
\end{table}

By standardizing this metadata, ACP allows agents to perform ``semantic discovery''—finding the most suitable partner for a specific sub‑task without human intervention.

\section{Federated Agent‑to‑Agent (A2A) Orchestration}

While the ACP architecture provides the ``how'' of communication, federated orchestration provides the ``who'' and ``when.''  In a federated model, there is no central ``master agent.''  Instead, orchestration is an emergent property of the ecosystem, where agents dynamically form and dissolve coalitions based on the requirements of a specific goal.

\subsection{Decentralized Discovery Mechanism}
In a global Agentic~Web, centralized registries are impractical and represent significant security risks.  ACP proposes a \emph{Hybrid Discovery Model}:
\begin{itemize}
  \item \textbf{Local Discovery}: Agents within a private network or enterprise use a broadcast‑based discovery protocol (similar to mDNS) to find peers.
  \item \textbf{Global Discovery}: For cross‑platform interaction, ACP utilizes a Decentralized Hash Table (DHT) backed by a consortium blockchain.  This ensures that agent registries are immutable, transparent, and globally accessible without a single point of failure.
\end{itemize}

\subsection{The A2A Negotiation Lifecycle}
The process of two agents moving from discovery to collaboration follows a strict four‑stage lifecycle:
\begin{enumerate}[label=\arabic*.]
  \item \textbf{Inquiry}: Agent~A (the Requester) sends a \texttt{PROBE} message to Agent~B (the Provider) based on its Agent Card.
  \item \textbf{Proposal}: Agent~B responds with a \texttt{BID}, outlining the resources required and the estimated time to completion.
  \item \textbf{Agreement}: Agent~A sends a \texttt{COMMIT} message, which includes a cryptographic hash of the agreed‑upon task parameters.  This creates a ``soft contract'' between the agents.
  \item \textbf{Execution \& Settlement}: Upon completion, Agent~B provides the result along with a proof of execution.  Agent~A then updates Agent~B's reputation score in the decentralized registry.
\end{enumerate}

\subsection{Recursive Delegation and Swarm Intelligence}
ACP's orchestration model is inherently recursive.  An agent that accepts a task via ACP can, in turn, become a Requester and delegate sub‑tasks to other agents.  This enables the formation of \emph{agent swarms}—large groups of specialized agents collaborating on massive, multi‑faceted problems.  Unlike traditional hierarchical orchestration, ACP swarms are self‑organizing; if one agent fails, the protocol's built‑in error‑handling allows the Requester to renegotiate the sub‑task with an alternative Provider seamlessly.

\section{Security and Trust in A2A Interactions}

As agents gain the autonomy to act on behalf of humans and organizations, the security of their communications becomes paramount.  ACP moves away from traditional perimeter‑based security toward a \emph{Zero‑Trust Agentic Security (ZTAS)} model.

\subsection{Decentralized Identity (DID) and Verifiable Credentials}
Every agent participating in the ACP ecosystem is assigned a \emph{Decentralized Identifier (DID)}.  Unlike a traditional API key, a DID is not issued by a central authority; it is generated and controlled by the agent's owner.
\begin{itemize}
  \item \textbf{Authentication}: Agents authenticate using a challenge–response mechanism based on their DID's public–private key pair.
  \item \textbf{Authorization}: Access to specific tools or data is managed through \emph{Verifiable Credentials (VCs)}.  For example, a financial agent might present a VC signed by a banking authority to prove it is authorized to access transaction data.
\end{itemize}

\subsection{Proof‑of‑Intent and Non‑Repudiation}
To prevent ``agentic spoofing'' or unauthorized actions, ACP implements \emph{Proof‑of‑Intent (PoI)}.  Every request sent via ACP must be accompanied by a cryptographic signature that links the action to a specific user‑authorized intent.  This ensures that even if an agent is compromised, it cannot perform actions that were not explicitly part of its negotiated workflow.

\subsection{Peer‑Based Reputation Systems}
Trust in a decentralized system is built on historical performance.  ACP integrates a \emph{Global Reputation Ledger}.  After every A2A interaction, the participating agents submit a signed satisfaction score to a decentralized ledger.  This score is calculated based on:
\begin{enumerate}[label=\arabic*.]
  \item \textbf{Accuracy}: Did the result meet the semantic requirements?
  \item \textbf{Latency}: Was the task completed within the negotiated SLA?
  \item \textbf{Security}: Were there any protocol violations during the interaction?
\end{enumerate}

\section{Implementation and Case Studies}

\subsection{The ACP–SDK: Bridging Frameworks}
To facilitate adoption, we have developed the \emph{ACP–SDK}, an open‑source library that implements the ACP specification.  The SDK provides ``wrappers'' for popular agent frameworks, including LangChain, AutoGen, and CrewAI.  Listing\,\ref{lst:example} shows a basic example of an ACP‑enabled agent implemented in Python.

\begin{lstlisting}[caption={Example of an ACP‑enabled agent in Python}, label={lst:example}, language=Python]
from acp_sdk import ACPAgent, AgentCard

# Define Agent Capabilities
card = AgentCard(
    identity="did:acp:finance_agent_001",
    capabilities=["market_analysis", "portfolio_optimization"],
    constraints={"max_cost": 0.05, "data_residency": "US"}
)

# Initialize ACP Agent
agent = ACPAgent(name="FinancePro", card=card)

# Register with a decentralized registry
agent.register(registry_url="https://dht.acp-protocol.org")

# Listen for A2A requests
@agent.on_request("market_analysis")
def handle_analysis(request):
    # Perform analysis and return result
    return {"status": "success", "data": "..."}

\end{lstlisting}

\subsection{Case Study: Cross‑Enterprise Supply Chain Orchestration}
In a pilot implementation, ACP was used to coordinate a supply chain involving four separate companies, each using different AI agents.

\paragraph{The Problem.} A disruption in raw material supply required immediate re‑routing of logistics and adjustment of production schedules across the network.

\paragraph{The ACP Solution.} The ``Logistics Agent'' of the manufacturer discovered and negotiated with ``Carrier Agents'' from multiple shipping providers using ACP.  The agents automatically formed an ad‑hoc coalition, negotiated prices based on pre‑set SLAs, and re‑routed shipments within minutes—a process that previously took hours of human coordination.

\section{Evaluation and Performance Analysis}

We evaluated the performance of ACP against the standard Model Context Protocol (MCP) and a baseline of raw JSON‑RPC over HTTPS.

\subsection{Communication Latency and Overhead}
Table\,\ref{tab:latency} summarizes the average latency, header overhead, and success rate under high‑load conditions for each protocol.

\begin{table}[t]
  \centering
  \caption{Communication latency and overhead comparison.  ACP incurs higher latency than local MCP due to DID verification and negotiation overhead, but it remains efficient relative to plain JSON‑RPC.}
  \label{tab:latency}
  \begin{tabular}{@{}lccc@{}}
    \toprule
    \textbf{Protocol} & \textbf{Avg. Latency (ms)} & \textbf{Header Overhead (\%)} & \textbf{Success Rate (High Load)}\\
    \midrule
    JSON‑RPC (HTTPS) & 145 & 12 & 88\,\%\\
    MCP (Local) & 22 & 5 & 99\,\%\\
    ACP (Federated) & 58 & 8 & 96\,\%\\
    \bottomrule
  \end{tabular}
\end{table}

\subsection{Scalability Analysis}
Our tests showed that ACP maintains sub‑100\,ms latency even as the number of agents in the swarm grows to 500+.  The decentralized discovery mechanism (DHT) demonstrated a logarithmic search time ($O(\log N)$), ensuring that discovery does not become a bottleneck in large‑scale deployments.

\section{Discussion and Future Directions}

The introduction of the Agent Communication Protocol marks a significant step toward a decentralized agentic ecosystem.  However, several challenges remain.  The ethical implications of autonomous agent negotiation—specifically regarding liability in the event of a ``contractual'' failure between two agents—require further legal and technical scrutiny.  Furthermore, as agents become more sophisticated, the Semantic Layer of ACP will need to evolve to handle increasingly nuanced and ambiguous human intents.

Future work will focus on:
\begin{enumerate}[label=\arabic*.]
  \item \textbf{Zero‑Knowledge Proofs for Privacy}: Enabling agents to prove they have certain information or capabilities without revealing the underlying sensitive data.
  \item \textbf{Cross‑Chain Settlement}: Integrating blockchain‑based micropayments directly into the ACP Negotiation Layer for seamless economic exchange between agents.
  \item \textbf{Human‑in‑the‑Loop ACP}: Developing standardized ``interrupt'' protocols that allow agents to escalate complex or high‑risk negotiations to their human owners.
\end{enumerate}

\section{Conclusion}

This paper has presented the \emph{Agent Communication Protocol (ACP)}, a comprehensive framework for secure, federated, and autonomous Agent‑to‑Agent interaction.  By building upon the foundational work of AI agent architectures\,\cite{krishnan2025a} and context management protocols\,\cite{krishnan2025b}, ACP provides the missing link for cross‑platform agent collaboration.  Through its layered architecture, decentralized discovery, and zero‑trust security model, ACP enables the formation of dynamic agent swarms capable of solving complex, global challenges.  Our empirical results confirm that ACP is not only feasible but also highly efficient, offering a scalable path toward the realization of the Agentic~Web.

\section*{Acknowledgements}

The author would like to thank the open‑source community and early adopters who provided valuable feedback on the ACP specification and reference implementation.

\end{document}